\documentclass[amsmath,amssymb,pra,final,showpacs,twocolumn]{revtex4-1}

\usepackage{bm,hyphenat,xspace}
\usepackage{graphicx,epsfig}

\newcommand{\scl}{0.64}

\newcommand {\mcu}{\mathcal{U}}

\begin{document}

\title {Universality in bosonic dimer-dimer scattering}

\author{A. Deltuva} 
\affiliation{Centro de F\'{\i}sica Nuclear da Universidade de Lisboa, 
P-1649-003 Lisboa, Portugal }

\received{May 16, 2011}

\begin{abstract}
Bosonic dimer-dimer scattering is studied near the unitary limit
using momentum-space equations for the four-particle transition operators. 
The impact of the Efimov effect on the dimer-dimer scattering
observables is explored and a number of universal relations 
is established with high accuracy. The rate for the creation of
Efimov trimers via dimer-dimer collisions is calculated.
\end{abstract}

 \maketitle

\section{Introduction}

Few-body systems with large two-particle scattering length $a$ 
possess a number of universal properties
that are independent of the short-range interaction details.
The three-particle system is simple enough such that 
analytic or semi-analytic derivation (at least of some) of those properties
is possible  \cite{braaten:rev}. Probably
the best-known example is the Efimov effect, the 
existence of an infinite number of weakly bound trimers with geometric
spectrum in the unitary limit $a \to \infty$ \cite{efimov:plb}.
The four-particle system, however, is too complicated for 
analytic approaches and therefore so far has been investigated mostly 
 numerically. 
While the bound state properties have been  calculated by several methods 
\cite{blume:00a,platter:04a,lazauskas:he,hammer:07a,stecher:09a},
the description of the four-particle continuum still constitutes a serious 
technical challenge, especially in the universal regime with many open channels;
first studies have been made using the hyperspherical formalism
\cite{dincao:09b,PhysRevLett.102.133201}.
Given the success of the four-nucleon scattering calculations
\cite{deltuva:07b,deltuva:08a} based on exact Alt, Grassberger, and Sandhas 
(AGS) equations \cite{grassberger:67} that were solved in momentum-space,
we applied them recently also to  bosonic atom-trimer 
scattering \cite{deltuva:10c,deltuva:11a} 
and established a number of universal relations between observables. 
In the present work we extend the technique of Ref.~\cite{deltuva:10c}
to study the universal physics in the bosonic dimer-dimer scattering;
in our model only one (shallow) dimer state exists.
We note that this problem has already been considered in 
Ref.~\cite{dincao:09a} using the adiabatic hyperspherical framework.
However, the AGS method enables us to determine the universal properties
with much  higher accuracy; in particular cases there are significant 
corrections to the results of Ref.~\cite{dincao:09a}.
Furthermore, we also present more detailed analysis
of dimer-dimer scattering observables.

\section{Dimer-dimer scattering equations}

A rigorous description of the four-particle scattering can be given
by the Faddeev-Yakubovsky equations \cite{yakubovsky:67}
for the wave-function components or by the equivalent 
 AGS integral equations \cite{grassberger:67} for the transition operators.
The latter ones are better suitable for the multichannel scattering
problem where the on-shell momenta and binding energies of the 
asymptotic states differ by many orders of magnitude \cite{deltuva:ef}.
Furthermore, the AGS equations lead more directly to the observables
since the on-shell matrix elements of the transition operators
calculated between the components of the corresponding initial and final 
channel states yield scattering amplitudes \cite{deltuva:07c}.
For the system of four identical bosons
we use the symmetrized form of the AGS equations. 
In the case of the atom-trimer scattering they
are given in Ref.~\cite{deltuva:10c}, whereas for
the dimer-dimer scattering they read
\begin{subequations} \label{eq:U}
\begin{align}  
\mcu_{12}  = {}&  (G_0  t  G_0)^{-1}  
 + P_{34}  U_1 G_0  t G_0  \mcu_{12} + U_2 G_0  t G_0  \mcu_{22} , 
\label{eq:U11} \\  
\mcu_{22}  = {}& (1 + P_{34}) U_1 G_0  t  G_0  \mcu_{12} . \label{eq:U21}
\end{align}
\end{subequations}
The transition operator $\mcu_{22}$ describes the elastic
scattering and $\mcu_{1 2}$ the transfer reactions leading to
the final atom-trimer states.
The dynamic input is the two-boson potential $v$. It yields
the two-boson transition matrix 
\begin{gather}
t = v+ v G_0 t 
\end{gather}
and the symmetrized transition operators
\begin{equation} \label{eq:U3}
U_{\alpha} =  P_\alpha G_0^{-1} + P_\alpha  t G_0  U_{\alpha}
\end{equation}
with $\alpha=1$  and 2 for the $1+3$   and $2+2$  subsystems, respectively.
In the above equations
$G_0 = (E+i0-H_0)^{-1}$ is the free resolvent for the
four-boson system with energy $E$ and kinetic energy operator $H_0$.
The permutation operators $P_{ij}$ of particles $i$ and $j$ and their
combinations $P_1 =  P_{12}\, P_{23} + P_{13}\, P_{23}$ and
$P_2 =  P_{13}\, P_{24} $ are sufficient to ensure the correct symmetry  
of the four-boson system when a special choice of the basis states is used
 \cite{deltuva:07a,deltuva:ef}. Namely,
the basis states must be symmetric under exchange
of particles 1 and 2; in addition,  for the 2+2 subsystem they
must be symmetric also under exchange of particles 3 and 4.

We employ momentum-space partial-wave framework \cite{deltuva:07a}
to solve the AGS equations \eqref{eq:U}.
After the discretization of momentum variables 
the system of integral equations
becomes a system of linear algebraic equations whose 
dimension is significantly reduced by using a
separable two-boson potential 
$v = |g\rangle \lambda \langle g| $
of rank 1 acting in the $S$ wave only.
Nevertheless, higher angular momentum states with $l_y, l_z \le 3$
(see  Ref.~\cite{deltuva:07a} for the notation) are taken into account
for the relative motion in $1+2$, $1+3$, and $2+2$ subsystems
such that the results are converged to at least four-digit accuracy
with respect to the partial-wave expansion.

To prove that our results are indeed independent of the short-range 
details of the interaction, 
we use two different choices of the potential  form factor
$\langle k |g\rangle = [1+c_2\,(k/\Lambda)^2]e^{-(k/\Lambda)^2}$
with $c_2 = 0$ and  $c_2=-9.17$.
It was shown in Ref.~\cite{deltuva:10c} that they  yield a very 
different trimer ground state and related observables, 
but, nevertheless, lead to the
same universal relations for the atom collisions with highly excited trimers
where the finite-range corrections become negligible.

\section{Results}

We study the  dimer-dimer scattering observables as functions
of the atom-atom scattering length $a$. 
We fix the geometry of the 
potential, i.e., the cutoff parameter $\Lambda$, whereas 
the strength $\lambda$ is adjusted to reproduce the desired value of $a$.
We present the results as dimensionless ratios that are independent
of  $\Lambda$ and particle mass $m$.

\begin{figure}[!]
\includegraphics[scale=0.45]{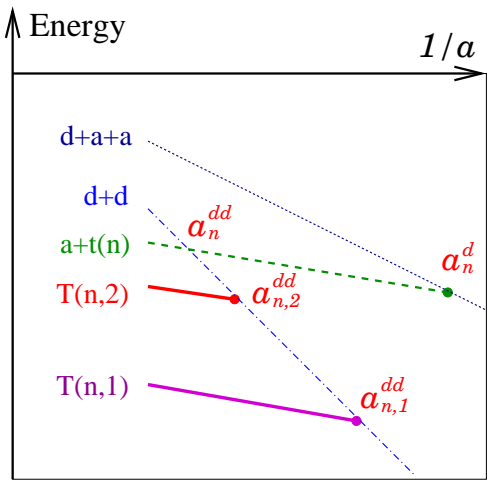} 
\caption{\label{fig:a} 
Schematic representation of the four-boson energy spectrum;
only one family ($n$) of multimers is shown.
The tetramer (T) energies, atom-trimer (a+t),  dimer-dimer (d+d),
and dimer-atom-atom (d+a+a) thresholds are displayed as  
solid, dashed, dashed-dotted, and dotted lines,
respectively. They intersect at special values of $a$ described in the text.
For a better visualization only qualitative relations are preserved.}
\end{figure}

A schematic $a$-dependence of the energy levels and thresholds
in the four-boson system is shown in Fig.~\ref{fig:a}.
As a reference point for $a$ we choose the intersection of the dimer-dimer and 
$n$th atom-trimer thresholds, i.e., $b_n = 2 b_d$ at $a = a_n^{dd}$,
where $b_d$ and $b_n$ are the binding energies of the 
dimer and the $n$th trimer, respectively.
A point important for three-boson physics is $a = a_n^{d}$ where
the $n$th trimer is at  the atom-dimer threshold, 
i.e.,  $b_n = b_d$.
Furthermore, the existence of two tetramers ($k=1,2$)
associated with each Efimov trimer
was predicted in Refs.~\cite{hammer:07a,stecher:09a}.
Except for the lowest two, all other tetramers lie
above the trimer ground state ($n=0$) and therefore are 
unstable bound states \cite{deltuva:10c,deltuva:11a} with finite width
$\Gamma_{n,k}$; their position relative to the four free particle threshold
is $-B_{n,k}$. At  $a = a_{n,k}^{dd}$ the corresponding tetramer
intersects the dimer-dimer threshold, i.e., 
 $B_{n,k} \approx 2b_d$, $\Gamma_{n,k} = 0$, leading to resonant effects
in the low-energy dimer-dimer scattering. 
The relations between these special $a$ values are collected in 
Table~\ref{tab:a} for $n=3$ and 4; to achieve the universal limit
with good accuracy we concentrate on the regime with $n$ and $a$ large enough,
$a\Lambda/2$ ranging from $10^3$ to $10^6$.
The  differences between the results obtained with $c_2 = 0$ and  
$c_2=-9.17$ in the potential  form factor 
are below 0.3\% in the  $n=3$ case and below 0.015\% in the  $n=4$ case.
 In addition we include  $ a_{n}^{d} \sqrt{m b_n^u}$ where
$b_n^u$ is the $n$th trimer binding energy in the unitary limit.
The agreement between our predictions and the semi-analytical result 
$\lim_{n \to \infty} a_{n}^{d} \sqrt{m b_n^u} \approx 0.0707645 $
given in Ref.~\cite{braaten:rev}  is of the same quality as for 
other ratios in Table~\ref{tab:a}, i.e.,
it is better than  0.015\% in the $n=4$ case.
We therefore conclude that our $n=4$ calculations provide the 
 universal results for dimer-dimer collisions with high accuracy,
\begin{subequations} \label{eq:ar-n}
\begin{align}
a_{n,1}^{dd}/ a_{n}^{dd} & =  0.3235(1), \\
a_{n,2}^{dd}/ a_{n}^{dd} & =  0.99947(2), \\
a_{n}^{dd}/ a_{n}^{d} & =  6.789(1);
\end{align}
\end{subequations}
the errors are estimated from the residual dependence on
$c_2$ and included $l_y, l_z$.
The above ratios calculated in  Ref.~\cite{dincao:09a}
with local potential at $n=2$
are 0.352, 0.981, and 6.73, respectively. Thus, they differ
from our results by 1 to 9\%. In particular, our $1 - a_{n,2}^{dd}/ a_{n}^{dd}$
value is smaller by a factor of 35, indicating that the shallow tetramer 
  is much closer to the corresponding 
atom-trimer threshold and thereby much stronger
affects the atom-trimer low-energy scattering  \cite{deltuva:11a}.

\begin{table}[!]
\begin{ruledtabular}
\begin{tabular}{*{5}{l}} $n$ & 
$ a_{n,1}^{dd}/ a_{n}^{dd}$ & $ a_{n,2}^{dd}/ a_{n}^{dd}$ & 
$ a_{n}^{dd}/ a_{n}^{d}$ & $ a_{n}^{d} \sqrt{m b_n^u}$ 
\\  \hline
3 & 0.32335 & 0.99948 & 6.8019 & 0.070601 \\
4 & 0.32352 & 0.99947 & 6.7896 & 0.070754 \\
\hline
3 & 0.32369 & 0.99948 & 6.7809 & 0.070857  \\
4 & 0.32354 & 0.99947 & 6.7886 & 0.070768  \\
\end{tabular}
\end{ruledtabular}
\caption{ \label{tab:a}
Special values of atom-atom scattering length 
calculated using potential form factor with  $c_2 = 0$ (top) and  
$c_2=-9.17$ (bottom).}
\end{table}

\begin{figure}[!]
\includegraphics[scale=\scl]{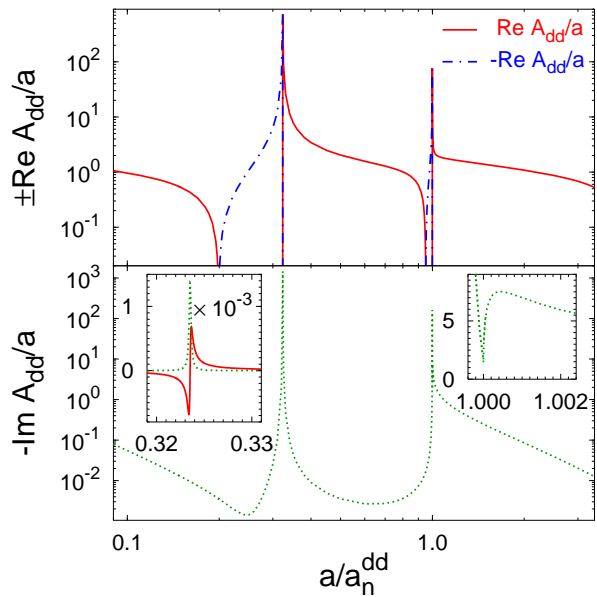}
\caption{\label{fig:add} (Color online)
Real and imaginary parts of the
dimer-dimer scattering length as functions of the atom-atom scattering length.
In the left inset  $\mathrm{Re}  A_{dd}/a$ 
 (solid curve) and  $\mathrm{Im}  A_{dd}/a$  (dotted curve) 
are shown in the vicinity of the $(n,1)$ tetramer
using linear scale. In the right inset $\mathrm{Im}  A_{dd}/a$ 
is shown in the vicinity of the dimer-dimer and atom-trimer threshold
intersection.}
\end{figure}

\begin{figure}[!]
\includegraphics[scale=\scl]{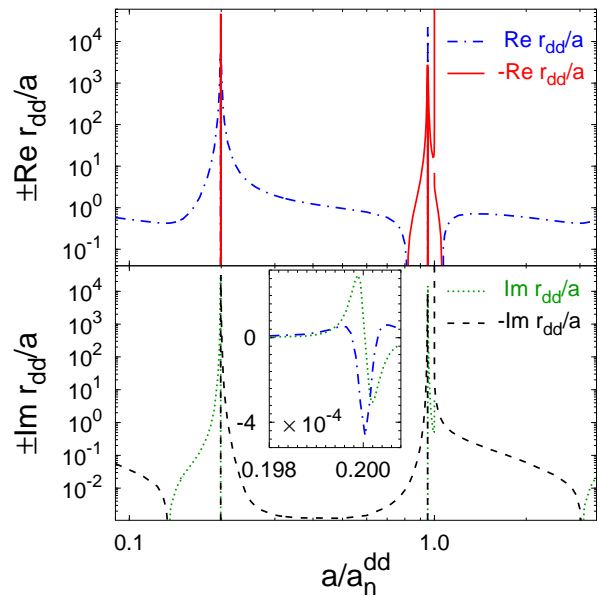}
\caption{\label{fig:rdd} (Color online)
Real and imaginary parts of the
dimer-dimer effective range parameter as functions of the atom-atom 
scattering length.
In the  inset  $\mathrm{Re} \, r_{dd}/a$ 
 (dash-dotted curve) and  $\mathrm{Im} \, r_{dd}/a$  (dotted curve) 
are shown in the vicinity of $\mathrm{Re} A_{dd} = 0$
using linear scale.}
\end{figure}

In the following we do not demonstrate explicitly
the convergence of our results,
however, they are checked to be independent of $c_2$ and
$n$ for $n \ge 3$ with good accuracy.
In  Figs.~\ref{fig:add} and \ref{fig:rdd}
we show the dimer-dimer scattering length $A_{dd}$
and the effective range parameter $r_{dd}$
as functions of $a$. In the universal limit
$A_{dd}/a$ and $r_{dd}/a$ are log-periodic  functions of $a$, i.e., the behavior
shown in Figs.~\ref{fig:add} and \ref{fig:rdd}
is repeated when $a$ increases by the
Efimov factor $e^{\pi/s_0} \approx 22.694$  \cite{braaten:rev}.
Since $A_{dd}/a$ and $r_{dd}/a$  vary over many orders of magnitude, 
we use logarithmic scale but distinguish between positive and negative values.
$A_{dd}$ exhibits a typical resonant behavior 
at $a= a_{n,k}^{dd}$; for $k=1$ it is
shown with finer resolution in the left inset of Fig.~\ref{fig:add}.
At the intersection of the dimer-dimer and atom-trimer thresholds
$A_{dd}$ has a cusp that is especially pronounced for the imaginary part
as demonstrated in the right inset of  Fig.~\ref{fig:add}.

The validity region of the effective range expansion is 
considerably narrower than the difference between the thresholds
$|2b_d - b_n|$. Thus,
the dimer-dimer effective range parameter $r_{dd}$ is ill-defined
at the intersection of the dimer-dimer and atom-trimer thresholds; 
 $\mathrm{Re} \, r_{dd}$  ($\mathrm{Im} \,r_{dd}$) diverges
when $a$ approaches   $a_{n}^{dd}$ from below (above)
as demonstrated in  Fig.~\ref{fig:rdd}. In contrast to $A_{dd}$,
$r_{dd}$ shows no resonant feature at $a= a_{n,k}^{dd}$ but 
exhibits very rapid variations
 when $\mathrm{Re} A_{dd}$ goes through zero,
i.e., at  $a/a_{n}^{dd} \approx 0.200$ and 0.951.
In a very narrow  region around these $a$ values
$\mathrm{Im} \,r_{dd}$ has positive and negative peaks whereas 
$\mathrm{Re} \,r_{dd}$ has two positive (negative) peaks 
separated by a much more pronounced negative (positive) peak
at  $a/a_{n}^{dd} \approx 0.200$ (0.951);
see the inset of Fig.~\ref{fig:rdd}. However, despite the dramatic 
variations of  $r_{dd}$, the quantity $|A_{dd}^2 r_{dd}|$ remains almost constant
in these regions; this is consistent with the corresponding behavior of the
two-particle system \cite{braaten:rev}.

\begin{figure*}[!]
\includegraphics[scale=\scl]{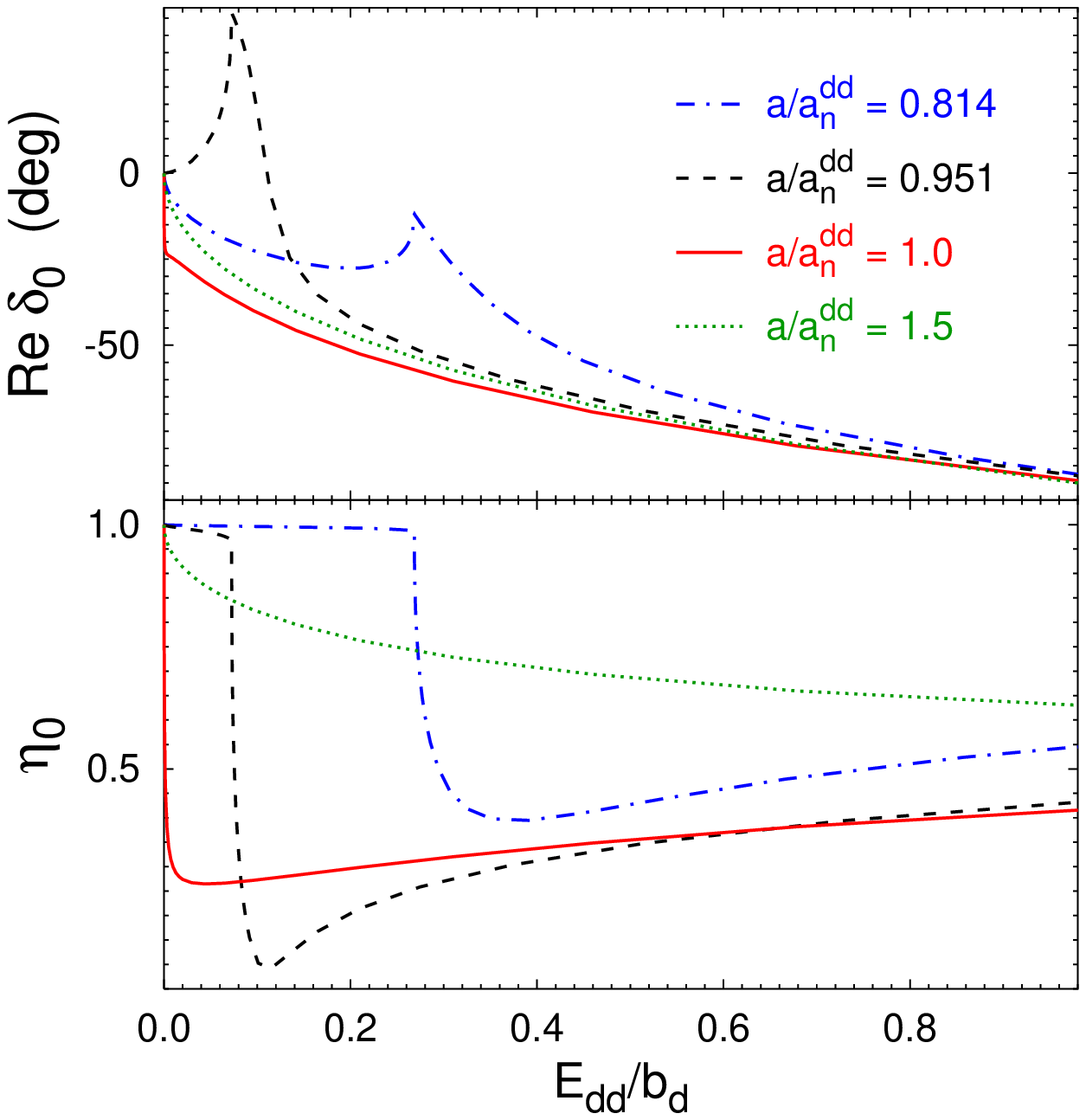} \quad\quad\quad
\includegraphics[scale=\scl]{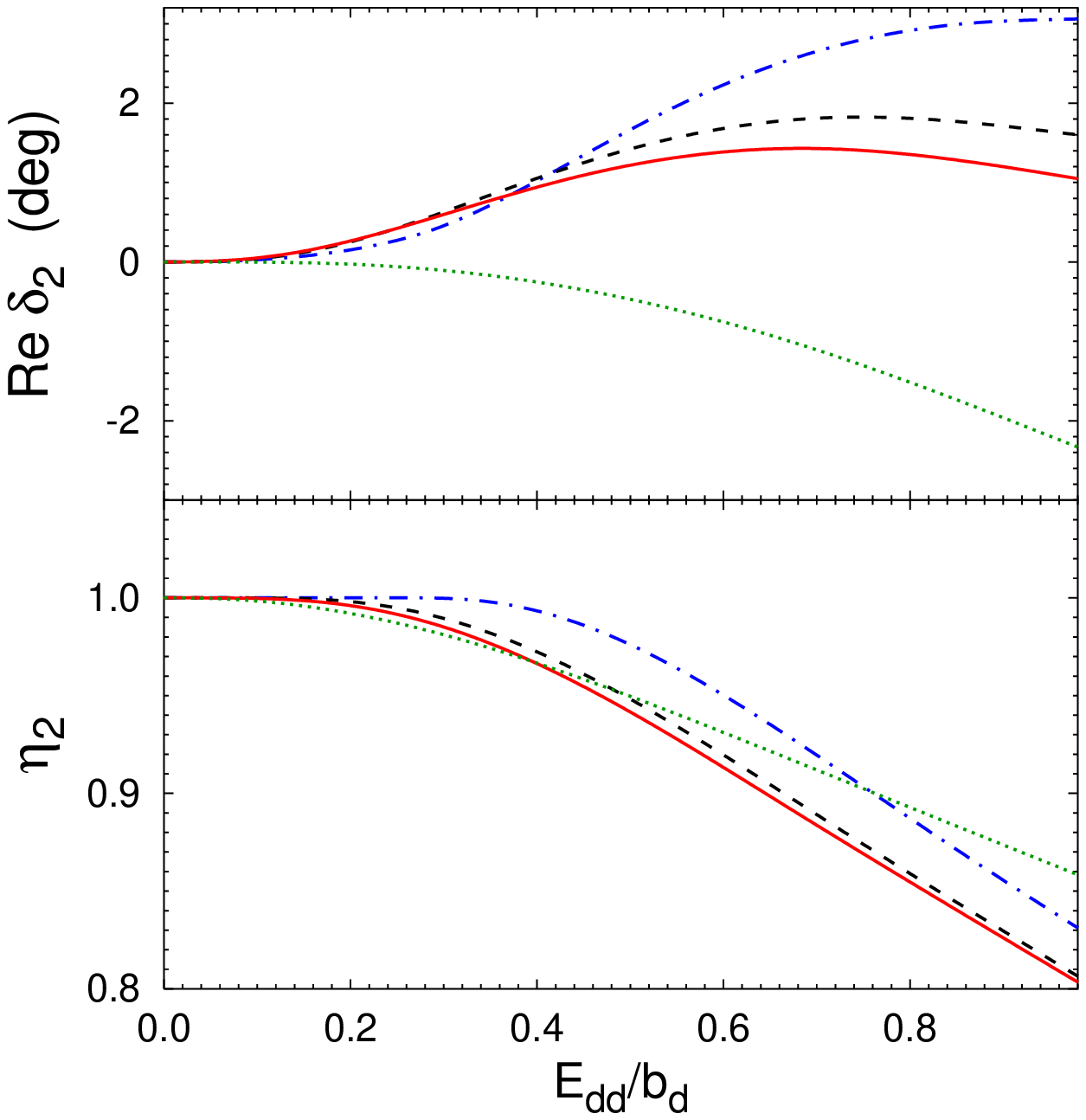} 
\caption{\label{fig:phase} (Color online)
Energy-dependence of the $S$- and $D$-wave
phase shifts and inelasticity parameters for the dimer-dimer scattering 
at specified values of  the atom-atom scattering length.
For a better comparison we set $\mathrm{Re} \, \delta_L =0$ at $E_{dd}=0$.}
\end{figure*}

Next we study the dimer-dimer  
complex phase shifts $\delta_L$ and inelasticity parameters 
$\eta_L = e^{-2\mathrm{Im} \,\delta_L}$ as functions of the 
relative dimer-dimer energy $E_{dd}$ up to the atom-atom-dimer threshold;
$L$ is the orbital angular momentum for the relative dimer-dimer motion.
 $S$-wave ($L=0$) results depend strongly on  $a$ as 
shown in Fig.~\ref{fig:phase} for few specific $a$ values.
We remind that the standard effective range expansion is not valid in the 
case of coinciding thresholds $a/a_{n}^{dd} = 1$.
A very rapid decrease of  $\mathrm{Re} \, \delta_0$ at  $a/a_{n}^{dd} = 1$ 
is caused by the proximity of the $(n,2)$ unstable tetramer.
At  $a/a_{n}^{dd} < 1$ the cusp in $\mathrm{Re} \, \delta_0$
and the descent of $\eta_0$ correspond to the opening 
of the $n$th atom-trimer channel. Due to centrifugal barrier these
features are absent in the $D$ wave ($L=2$).
$\mathrm{Re} \, \delta_2$ are small but, depending on $a$,
can be of both signs, while $\eta_2$ deviate from 1 quite significantly
as $E_{dd}$  increases. 
Thus, near $E_{dd} \approx b_d$ the $D$-wave contribution to the inelastic
cross section may exceed the one of the $S$ wave, the ratio being
$5(1-\eta_2^2)/(1-\eta_0^2)$. 
Higher waves can be neglected in the considered energy regime  $E_{dd} < b_d$, 
i.e., $\delta_L \approx 0$ for $L\ge 4$.

\begin{figure*}[!]
\includegraphics[scale=0.9]{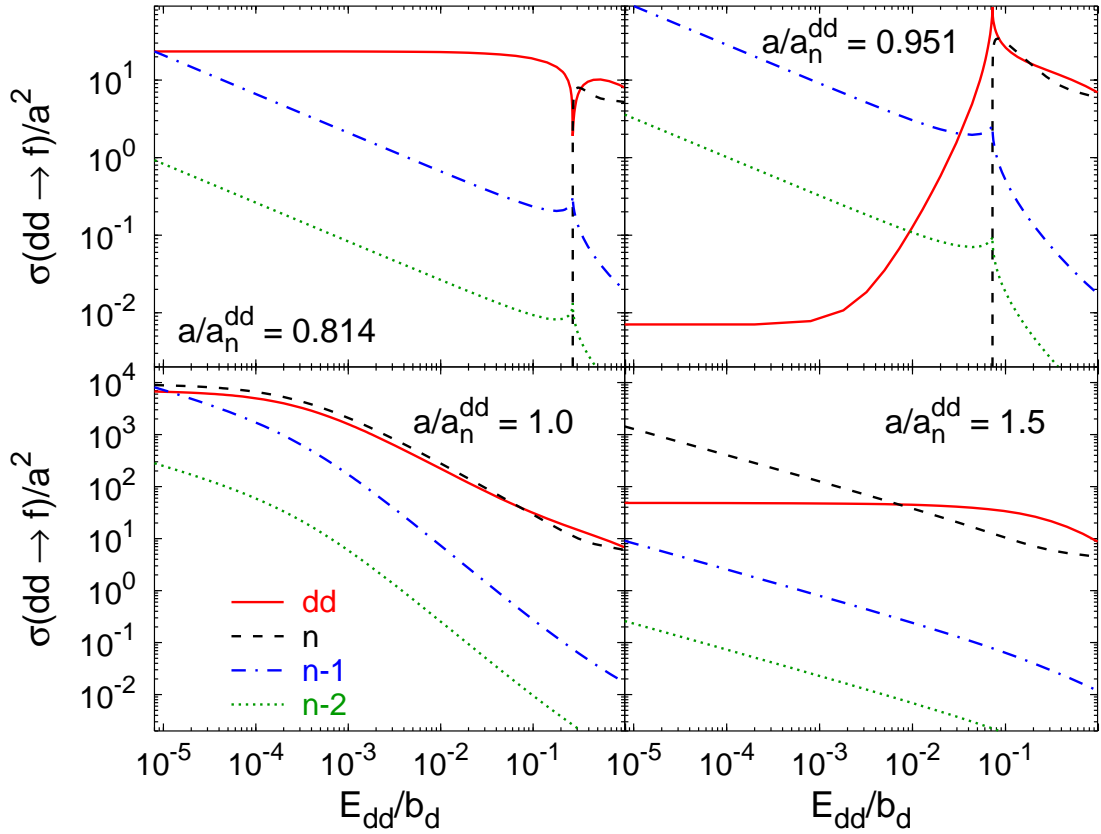} 
\caption{\label{fig:cs} (Color online)
Elastic and transfer cross sections 
for the dimer-dimer scattering 
at specified values of  the atom-atom scattering length.}
\end{figure*}

The cross sections for the dimer-dimer elastic scattering 
$\sigma(dd \to dd)$ and transfer/rearrangement reactions $\sigma(dd \to n')$
leading to an atom plus trimer in the $n'$th state
are shown  in Fig.~\ref{fig:cs}. 
At very low energy $\sigma(dd \to dd)$  become
nearly energy-independent (this is true also for  $a/a_{n}^{dd} = 1$
at  $E_{dd}/b_d < 10^{-5}$)
while  $\sigma(dd \to n')$ increase with decreasing $E_{dd}$
nearly as $E_{dd}^{-1/2}$. The exception is $\sigma(dd \to n)$
in the case of coinciding thresholds $a/a_{n}^{dd} = 1$ where
the increase slows down until $\sigma(dd \to n)$
becomes almost constant at   $E_{dd}/b_d < 10^{-5}$.
Thus, the very low energy behavior of the cross sections is
consistent with the Wigner threshold law.
The reason why for $a/a_{n}^{dd} = 1$ the region of
nearly constant  $\sigma(dd \to dd)$ and $\sigma(dd \to n)$ is
so narrow is the proximity of the $(n,2)$ unstable tetramer.
The inelastic scattering, apart from the very narrow regions just
above the atom-trimer thresholds,
 is dominated by the open channel with the shallowest trimer;
the reactions leading to more deeply bound trimers are strongly 
suppressed. 
The elastic cross section
 appears to be remarkably sensitive to the atom-atom scattering length.
For example,
$a/a_{n}^{dd} = 0.814$ corresponds to $\mathrm{Re} \, r_{dd} \approx 0$ 
leading to an almost constant $\sigma(dd \to dd)$ at $E_{dd}/b_d < 0.1$
while for $a/a_{n}^{dd} = 0.951$  where   $\mathrm{Re} \, A_{dd} \approx 0$ 
the elastic cross section is very small at $E_{dd}=0$ but 
then increases with energy, in contrast to the other shown cases.
Furthermore, the opening of the $n$th atom-trimer channel affects
the elastic dimer-dimer cross section
 in opposite ways for $a/a_{n}^{dd} = 0.814$  and 0.951.

\begin{figure}[!]
\includegraphics[scale=\scl]{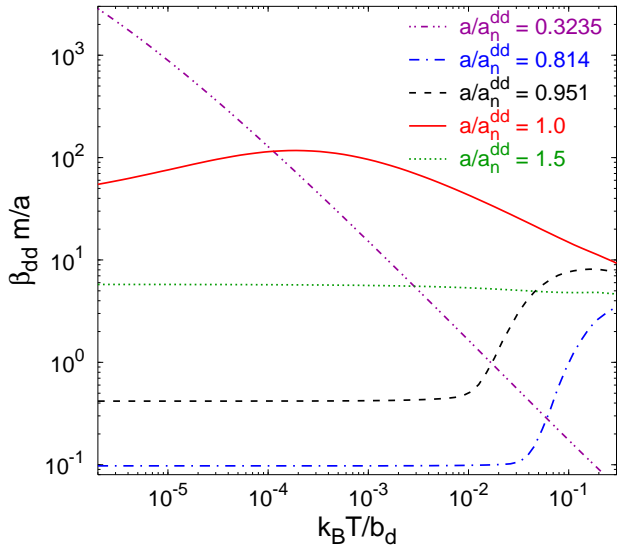} 
\caption{\label{fig:rlx} (Color online)
Temperature dependence of the trimer creation rate
at specified values of  the atom-atom scattering length;
$a/a_{n}^{dd} = 0.3235$ corresponds to the $(n,1)$st resonance
that leads to a peak of $\beta_{dd}$ at $T=0$.}
\end{figure}

The above results are indispensable for the description of the 
dimer-trimer conversion process, i.e.,
the creation of Efimov trimers via rearrangement reaction in a gas of dimers.
The total trimer creation rate in the dimer-dimer scattering 
$\beta_{dd} = \sum_{n'} \beta_{dd}^{n'}$
has contributions from  all open atom-trimer channels,  
\begin{equation} \label{eq:rlxb}
\beta_{dd}^{n'} = \langle v_{dd} \, \sigma(dd \to n') \rangle, 
\end{equation}
where $v_{dd} = \sqrt{2E_{dd}/m}$ is the relative dimer-dimer velocity
and $\langle \ldots \rangle$ denotes the thermal average.
At vanishing temperature the trimer creation rate
$ \beta_{dd}(0) = -(8\pi/m) \, \mathrm{Im} A_{dd} $
is determined by the results of  Fig.~\ref{fig:add};
thus, it gets resonantly enhanced around $a = a_{n,k}^{dd}$.
The results at finite temperature $T$,
assuming the Boltzmann distribution for the relative dimer-dimer energy,
 are presented in Fig.~\ref{fig:rlx} for few selected values of $a$.
The shown  examples differ remarkably in their $T$ dependence:
at very low temperatures
$\beta_{dd}$ decreases with $T$ very rapidly at the resonance
$(a/a_{n}^{dd} = 0.3235)$ 
 while in other cases  its variation is considerably slower.
A more pronounced increase of $\beta_{dd}$ at higher temperature
for $a/a_{n}^{dd} = 0.814$ and 0.951 is due to the opening of the 
$n$th atom-trimer channel.
We do not show predictions for individual $\beta_{dd}^{n'}$,
but the results of Fig.~\ref{fig:cs} clearly indicate that shallow trimers 
are produced most efficiently.

Finally we note that $\beta_{dd}$ was calculated also in Ref.~\cite{dincao:09a}
where it was called the dimer-dimer relaxation rate. 
There is a good qualitative agreement between our  $\beta_{dd}$
results and those of Ref.~\cite{dincao:09a}. There are, however,
some quantitative differences, e.g., at  $a/a_{n}^{dd} =1.5$ our
$\beta_{dd}$ decreases with $T$ slower. At least to some extent this is 
probably caused by the significant $D$-wave contribution taken into account 
in our calculations.

The only available experimental data \cite{ferlaino:08a} refer to
the regime where $a$ exceeds the interaction range by a factor 
1 to 10 and therefore  non-negligible finite-range corrections can be expected.
In this respect our calculations present no improvement over those
of Ref.~\cite{dincao:09a} that could provide only qualitative
explanation of some features of the data. However, 
our predictions would be valuable for the future
experiments performed in the universal regime.

\section{Summary}
We studied bosonic dimer-dimer scattering near the unitary limit.
It is a complicated multichannel four-particle scattering problem
involving, in the present calculations, one dimer-dimer and up to
five atom-trimer channels with very broad range of binding energies.
Exact AGS equations were solved using momentum-space techniques.
The log-periodic structure of the dimer-dimer scattering observables
was found to be independent of the short-range potential details.
Universal results for 
the dimer-dimer scattering length, effective range, phase shifts, elastic
and transfer cross sections, some tetramer properties, and trimer creation
rate were obtained with an accuracy considerably higher than in previous works.


\end{document}